\documentclass[12pt]{article}
  \usepackage{amsfonts}
\usepackage{amssymb}
\usepackage{amsmath}
\usepackage{hyperref}
\newcommand{\ead}[1]{\vspace*{5pt} {E-mail: \tt{#1}}}

\begin{document}
\title{\bf Exact Solutions in Nonlocal Linear Models}
\vspace{-1.27cm}
\author{S.~Yu.~Vernov\footnote{\ead{svernov@theory.sinp.msu.ru}}\\
Skobeltsyn Institute of Nuclear Physics, Moscow State
University,\\
Vorobyevy Gory, 119991, Moscow, Russia}
\date{}
\maketitle \vspace{-7.2mm}
\begin{abstract}
A general class of cosmological models driven by a nonlocal scalar
field inspired by the string field theory is studied. Using the fact
that the considering linear nonlocal model is equivalent to an
infinite number of local models we have found an exact special
solution of the nonlocal Friedmann equations. This solution
describes a monotonically increasing Universe with the phantom dark
energy.
\end{abstract}

\section{Introduction}

Recently string theory has been intensively discussed as promising
candidates for the theoretical explanation (see for example
\cite{string-cosmo,Biswas}) of the obtained experimental
data~\cite{cosmo-obser}.

The purpose of this paper is to present recent results concerning
studies of  the string field theory (SFT) inspired nonlocal
cosmological models in the Friedmann--Robertson--Walker Universe.
A Distinguished feature of these
models~\cite{IA1}--\cite{Jukovskaya0707} is the presence of an
infinite number of higher derivative terms. For special values of
the parameters these models describe linear approximations to the
cubic bosonic or nonBPS fermionic SFT nonlocal tachyon models,
p-adic string models or the models with the invariance of the
action under the shift of the dilaton field to a constant. The
NonBPS fermionic SFT nonlocal tachyon model has been considered as
a candidate for the dark energy~\cite{IA1}. This report is based
on papers~\cite{AJV0701184}, which have been made in cooperation with I.Ya.~Aref'eva and
L.V.~Joukovskaya.

\section{Nonlocal linear models}
Let us consider a model of gravity coupling with a SFT inspired
nonlocal scalar field
 \begin{equation}
\label{action-1}  S=\int
d^4x\sqrt{-g}\left(\frac{m_p^2}{2}R+\frac{\xi^2 }{2}
\phi\,\square_g\phi+
 \frac{1}{2}\left(\phi^2-c\,\Phi^2\right)-\Lambda^\prime\right),
\end{equation}
where $\Phi=e^{\square_g}\phi$,  $g_{\mu\nu}$ is the metric tensor
(the signature is $(-,+,+,+)$ ),
$\square_g=\frac1{\sqrt{-g}}\partial_{\mu}\sqrt{-g}g^{\mu\nu}\partial_{\nu}$,
 $m_p^2=g_4M_p^2/M_s^2$, $M_p$ is  a mass Planck,
 $M_s$ is a characteristic string scale
 related with the  string tension $ \alpha^{\prime}$:
$M_s=1/\sqrt{ \alpha^{\prime}}$, $\phi$ is a dimensionless scalar
field and $g_4$ is a dimensionless four dimensional effective
coupling constant. An effective four dimensional cosmological
constant is  $\Lambda=\frac{M_s^4}{g_4}\Lambda^\prime $.
Parameters $\xi$ and $c$ are positive. The term
$(e^{\square_g}\phi)^2$
 is analogous to the
interaction term for the tachyon in the string action.

We consider the case of the spatially flat
Friedmann--Robertson--Walker Universe:
\begin{equation}
\label{mFr} ds^2={}-dt^2+a^2(t)\left(dx_1^2+dx_2^2+dx_3^2\right)
\end{equation}
and spatially homogeneous solutions $\phi(t)$.  In this case
$T_{\alpha\beta}=g_{\alpha\beta}\,\mbox{diag}\{{\cal E},{-\cal
P},{-\cal P},{-\cal P}\}$,  where the energy density ${\cal E}$
and pressure ${\cal P}$ are as follows
\begin{equation}
\label{E-phi-H-p}
 {\cal E}={\cal E}_k+{\cal E}_{nl2}+{\cal E}_p+{\cal E}_{nl1}+\Lambda^\prime,\
\qquad {\cal P}={\cal E}_k+{\cal E}_{nl2}-{\cal E}_p-{\cal
E}_{nl1}-\Lambda^\prime,
\end{equation}
\begin{equation}
\label{El-Enl}
\begin{array}{l}
\displaystyle {\cal E}_k=\displaystyle
\frac{\xi^2}{2}(\partial_0\phi)^2,\qquad {\cal
 E}_{nl2}={} -c\int\limits_{0}^{1}\left(\partial
e^{(1+\rho){\cal D}} \phi\right) \left(\partial e^{(1-\rho) {\cal
D}}\phi\right) d \rho,\\[2.7mm]
\displaystyle {\cal E}_p={}
 -\frac{1}{2}\left(\phi^2-c\,\left(e^{\cal D}\phi\right)^{2}\right),\qquad
{\cal E}_{nl1}= c\int\limits_{0}^{1}\left(e^{(1+\rho) {\cal D}}
\phi\right)
 \left(-{\cal D} e^{(1-\rho){\cal D}}  \phi\right) d \rho,
\end{array}
\end{equation}
where ${\cal D}\equiv {}- \partial_0^2- 3H(t)\partial_0$, the
Hubble parameter $H\equiv \dot{a}/a$ and dot denotes the time
derivative ($\dot a\equiv\partial_0 a$). Note that the
energy-momentum tensor $T_{\alpha\beta}$ includes the nonlocal
terms, so the Einstein's equations
\begin{equation}
3H^2= \frac{1}{m_p^2}~{\cal E}, \qquad \dot
H={}-\frac{1}{2m_p^2}~({\cal E}+{\cal P}) \label{eomprho}
\end{equation}
are nonlocal ones. The second equation of (\ref{eomprho}) is the
nonlinear integral equation in $H(t)$:
\begin{equation}
\dot H={}-\frac{1}{m_p^2}\left(
\frac{\xi^2}{2}(\partial_0\phi)^2-c\int\limits_{0}^{1}\!\left(\partial_0
e^{(1+\rho){\cal D}} \phi\right) \left(\partial_0 e^{(1-\rho)
{\cal D}}\phi\right)d\rho\right).
\end{equation}

\section{The equation of motion}

In the spatially flat Friedmann--Robertson--Walker Universe we get
the following equation of motion for the space homogeneous scalar
field $\phi$
\begin{equation}
\label{equphiNFrid} (\xi^2{\cal D} +1) e^{-2 {\cal D}}\phi=
c\,\phi.
\end{equation}
Really this equation is a consequence of the Einstein's
equations, hence, both the metric $g_{\mu\nu}$ and the scalar
field $\phi$ are unknown. We assume that the metric $g_{\mu\nu}$
is given and consider eq.~(\ref{equphiNFrid}) as an equation in
$\phi$. Solutions of the following linear differential equation
\begin{equation}
\label{equ1} -{\cal
D}\phi\equiv\ddot\phi(t)+3H(t)\dot\phi(t)=\alpha^2\phi,
\end{equation}
 represent the solutions of eq.~(\ref{equphiNFrid})
 with $\alpha$, which is a root of the
characteristic equation
\begin{equation}
F(\alpha^2)\equiv-\xi^2\alpha^2  + 1 - c\:e^{-2\alpha^2}=0.
\label{5c}
\end{equation}
Equation (\ref{5c}) has the following roots

\begin{equation}
\label{sol-lamb-k} \alpha_n={}\pm\frac{1}{2\xi}\sqrt{4+ 2\xi^2
W_n\left({}-\frac{2c\:e^{-2/\xi^2}}{\xi^2}\right)}, \quad n=0,\pm
1,\pm 2, \dots,
\end{equation}
where $W_n$ is  the n-s branch of the  Lambert function,
satisfying a relation $W(z)e^{W(z)}=z$. The Lambert function is a
multivalued function, so eq.~(\ref{5c}) has an infinite number of
roots. Parameters $\xi$ and $c$ are real, therefore if $\alpha_n$
is a root of (\ref{5c}), then the adjoined number $\alpha_n^*$ is
a root as well. Note that  if $\alpha_n$ is a root of (\ref{5c}),
then $-\alpha_n$ is a root as well.

 If $\alpha^2=\alpha^2_0$ is a multiple root, then at this point
$F(\alpha^2_0)=0$ and $F'(\alpha^2_0)=0$. Double roots exist if
and only if
\begin{equation}
\label{c-xi} c=\frac{\xi^2}{2}e^{(2/\xi^2-1)}.
\end{equation}

Note that existence of double roots means that there exist
solutions of equation~(\ref{equphiNFrid}), which do not satisfy
to equation (\ref{equ1}), but satisfy the following equation
$\square_g^2 \phi =\alpha^4\phi$. In the flat case an example of a
such solution is the function $\phi(t)=t\exp(\alpha t)$
(see\cite{AJV0701184}).
 All roots for any
$\xi$ and $c$ are no more than double degenerated, because
$F''(\alpha^2_0)\neq 0$. We consider such values of $\xi$ and $c$
that  equality (\ref{c-xi}) is not satisfied and all roots are
single.

Let us make an assumption, that $\phi_1(t)$ and $H_1(t)$ satisfy
eq.~(\ref{equ1}), with $\alpha=\alpha_1$ is a root of
eq.~(\ref{5c}), hence,  eq.~(\ref{equphiNFrid}) is solved. The
energy density and the pressure  are as follows:
\begin{equation*}
 E(\phi_1)=\frac{\eta_{\alpha_{1^{\vphantom {27}}}}}{2}
 \left(\left(\partial_0\phi_1\right)^2-
    \alpha_1^2\phi_1^2\right),
\qquad
 P(\phi_1)=\frac{\eta_{\alpha_{1^{\vphantom {27}}}}}{2}\left(
    \left(\partial_0\phi_1\right)^2+
    \alpha_1^2\phi_1^2\right),
\qquad     \eta_{\alpha_{1^{\vphantom {27}}}}\equiv
\xi^2+2\xi^2\alpha_1^2-2.
\end{equation*}

Using this formula, we rewrite system~(\ref{eomprho}) in the
following form:
\begin{equation}
3H_1^2= \frac{\eta_{\alpha_{1^{\vphantom
{27}}}}}{2m_p^2}\left(\dot\phi_1^2-
    \alpha_1^2\phi_1^2+\Lambda'\right),\qquad
\dot H_1={}-\frac{\eta_{\alpha_{1^{\vphantom
{27}}}}}{2m_p^2}\dot\phi_1^2.
 \label{eomprholocal1}
\end{equation}
So, our assumption allows to transform a system with a nonlocal
scalar field into a system with a local scalar field $\phi_1$.
Note that in such a way we obtain an infinity number of local
systems, because eq.~(\ref{5c}) has an infinity number of roots.

\section{Exact Solutions}

In this paper we present an exact solution, which looks realistic
for the SFT inspired cosmological model. One of the possible
scenarios of the Universe evolution considers the Universe to be a
D3-brane (3 spatial and 1 time variable) embedded in
higher-dimensional space-time. The D-brane is unstable and does
evolve to the stable state. This process is described by the open
string dynamics, which ends are attached to the brane. A phantom
scalar field is an open string tachyon. According to the
Sen's conjecture~\cite{Sen-g}, the tachyon describes brane decay, at which a
slow transition to the stable vacuum, correlating with only states of
the closed string, takes place. This picture allows us to specify
the asymptotic conditions for the phantom field $\phi(t)$. We
assume that $\phi(t)$ smoothly rolls from the unstable
perturbative vacuum ($\phi=0$) to a nonperturbative one, for
example $\phi=A_0$, where $A_0$ is a nonzero constant, and stops
there. In other words we seek a king-like solution.

 At $c=1$ one of solutions of eq.~(\ref{5c}) is $\alpha=0$ and
at $\Lambda^\prime>0$ we obtain a real solution
\begin{equation}
\label{Hnew}
H_1(t)=\sqrt{\frac{\Lambda^\prime}{6m_p^2}}\tanh\left(\,\tilde{t}\,\right),\quad
\phi_1(t)=\pm\sqrt{\frac{2m_p^2}{3(2-\xi^2)}}\arctan
\left(\sinh\left(\,\tilde{t}\, \right)\right)+C_1,
\end{equation}
where $\tilde{t}\equiv\sqrt{\frac{3\Lambda^\prime}
{2m_p^2}}(t-t_0)$, $t_0$ and $C_1$ are arbitrary real constants.
We have obtained that $\phi(t)$ can be a real scalar field if and
only if it is a phantom one ($\eta_\alpha<0$, that is equivalent
to $\xi^2<2$). We have obtained the general solution for
(\ref{eomprholocal1}). The Hubble parameter $H(t)$ is a
monotonically increasing function, so solution (\ref{Hnew})
corresponds to phantom dark energy.

\section{Conclusions}

We have studied the SFT inspired linear nonlocal model. This model
has an infinite number of higher derivative terms and are
characterized by two positive parameters. For some values of these
parameters the corresponding actions are linear approximations to
either the bosonic or nonBPS fermionic cubic SFT as well as to the
nonpolynomial SFT.

We have shown that our linear model with one nonlocal scalar field
generates an infinite number of local models. Some of these models
can be solved explicitly and, hence, special exact solutions for
nonlocal model in the Friedmann metric can be
obtained~\cite{AJV0701184,AVzeta}. We have constructed an exact
kink-like solution, which correspond to monotonically increasing
Universe with phantom dark energy. Note that the obtained
behaviour of the Hubble parameter is close to behavior of the
Hubble parameter in the nonlinear nonlocal model~\cite{IA1}, which
recently has been obtained numerically~\cite{Jukovskaya0707}. Note
that the considering nonlocal model generates local models with an
arbitrary number of local scalar fields as well~\cite{AJV0701184}.

Author is grateful to I.Ya. Aref'eva and L.V. Joukovskaya for the
collaboration and useful discussions. Author is thankful to the
organizers of {\href{http://theor.jinr.ru/~sqs07/}{the International Workshop on
Supersym\-metries}} \  {\href{http://theor.jinr.ru/~sqs07/}{and Quantum
Symmetries}} ("SQS'07") for hospitality and financial support. This
research is supported in part by RFBR grant 05-01-00758 and by
Russian President's grant NSh--8122.2006.2.

\end{document}